# Mechanism of Cooper-pairing in layered high temperature superconductors

A. Tavkhelidze

Ilia State University, 3-5 Cholokashvili Avenue, Tbilisi 0162, Georgia
avtandil.tavkhelidze@iliauni.edu.ge

In this study, the pairing mechanism for layered HTS materials based on attraction between electrons from adjacent layers is proposed. Initially, each layer has expanded Fermi sphere owing to ridged geometry. When the two layers are close enough for tunneling, it becomes energetically advantageous to form correlated quantum states (CQS), reducing the Fermi sphere volume. Cooper pairs, comprising inter-tunneling electrons, occupy the CQS. The image force is responsible for the electron-electron attraction. Pair-binding energy and the corresponding effective mass vary in a wide range. At T>0, some heavy pairs do not condense. Such pairs are responsible for pseudogap. Light pairs get Bose condensed and are responsible for superconductivity. The proposed mechanism provides clarification of superconductivity in cuprates, iron based superconductors and LSCO/LCO interfaces. It provides explanation of two energy gaps and two characteristic temperatures in layered superconducting materials. It also provides clarification on the Fermi surface pockets, anisotropy of charge transport in pseudogap state, and other properties of HTS materials. The pseudogap, estimated within the model, fits the experimental values for the two-layer cuprates, such as YBCO, Bi2212, Tl2212, and Hg1212.

**1. Introduction**

Iron based superconductors are the first non-cuprate materials exhibiting superconductivity at relatively high temperatures.[1] The 2D electronic structure, the superconducting dome in the phase diagram, the anomaly of transport properties in under doped regime etc., make iron based superconductors similar to cuprates. These materials have a crystal structure comparable with cuprates. Like cuprates, they have layered structure and non-planar geometry of the layers.

At the base of crystal structural similarities, and taking into account recently discovered unconventional properties of ridged layers, we suggest that Cooper pairing (in both types of materials) emerges from layer geometry. The presented mechanism is based only on the layered structure of the material and non-planar layer geometry. It is equally applicable to cuprates and iron based superconductors. Such a mechanism is also supported by recent experiments on interface superconductivity.[2] Interface of non-superconducting materials $La_2CuO_4$ and $La_{1.55}Sr_{0.45}CuO_4$ exhibit superconductivity. Both have layered structure and non planar geometry of the layers.

In a high temperature superconductor (HTS) materials, the Cooper pairs are carriers of the superconducting current. However, high critical temperature $T_c$, low-order parameter, and the unconventional isotopic effect indicate that the phonon mechanism of pairing is not applicable. In the HTS cuprates, two separate energy gaps exist.[3] Fermi surface pockets were found in quantum oscillations of hall coefficient [4, 5] and angle-resolved photoemission spectroscopy (ARPES).[6] The $CuO_2$ layers are responsible for superconductivity, and the electrons are concentrated in them. Reduction of number of $CuO_2$ layers in the ultra-thin films leads to decrease in $T_c$. Furthermore, superconductivity vanishes when less than two layers are left,[7] indicating that superconductivity emerges from some interlayer effect. Seemingly, this contradicts with the results obtained from interface superconductivity where, the single $CuO_2$ lawyer is responsible for superconductivity.[8] However, superconductivity emerges only in the presence of the interfacing material, providing another layer. In this study, the possible pairing mechanism, based on single-electron tunneling between $CuO_2$ layers, is proposed.

Recently, it has been found that the ridged thin films exhibit unconventional properties. Ridges impose additional boundary conditions on the electron wave function and some quantum states become forbidden. Rejected electrons occupy quantum states with higher energies. The Fermi vector, $k_F$, and Fermi energy, $E_F$, are increased in the ridged geometry,[9, 10] which can be termed as Fermi sphere expansion (FSE), for convenience. Pairing mechanism presented in this study is based on the assumption that $CuO_2$ layers, like ridged films, exhibit FSE. We divide cuprate material into $CuO_2$ layers, each containing electron gas modified by FSE. Subsequently, we consider the interaction of the adjacent layers, through single-electron tunneling. Cooper pairs exist in correlated quantum states (CQSs), and such states belong to the system of two or more $CuO_2$ layers. In our model pairs do not exist in one particular layer, as in the Lawrence–Doniach model [11] or the electron confinement model.[12] The electron–electron attraction originates from the image force.

The objectives of this study are to introduce a possible mechanism of Cooper-pair formation, calculate the pseudogap value on its base, and compare it with the experimental results. In Sec. 2 we describe general properties of ridged layers. In Sec. 3 we illustrate energy reduction in the system of two adjacent ridged layers, introduce CQS, describe single electron tunneling as a possible way of energy reduction, introduce an electron - electron attraction mechanism, and demonstrate that CQS can be occupied by the Cooper pairs. In Sec. 4 we apply the general results obtained for ridged layers to $CuO_2$ layers and estimate the layer binding energy. In Sec. 5 we estimate the binding energy per electron and compare it with experimental values for two layer cuprates. In Sec. 6 we look at Bose condensation in our model and introduce free (not condensed)

Cooper pairs to explain two energy gaps and two characteristic temperatures. In Sec. 7 we elucidate how Fermi surface pockets emerge from layer geometry. In Sec. 8 we try to clarify some unconventional experimental dependences using our model. The main conclusions of our study are given in Sec. 9.

**2. Ridged layer properties**
Figure 1 shows a reference quantum well layer (a) and a ridged quantum well (RQW) layer (b), and the corresponding energy diagrams. The ridges have depth $a$ and period $2w$. The thickness of reference well layer $L+a/2$ is chosen so that the two layers have the same volume (per unit area). Owing to ridges, some quantum states become forbidden in an RQW.

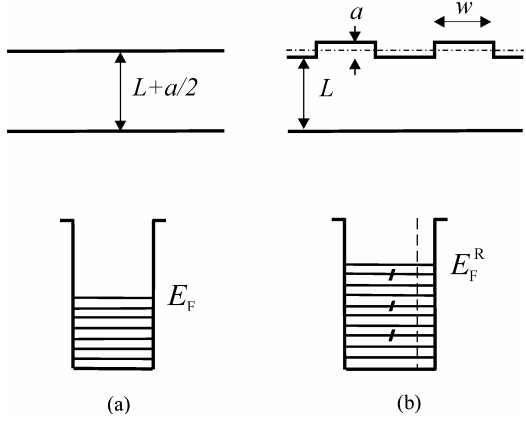

Fig. 1 (a) Reference quantum well layer and its energy diagram (b) RQW layer and its energy diagram. Horizontal lines depict energy levels (quantum states). Ridge forbidden energy levels are shown as crossed lines. For simplicity, we assume that the energy levels are equidistant and do not change position in RQW.

The rejected electrons have to occupy high energy levels, and the Fermi energy increases from $E_F$ to $E_F^{(R)}$.[13] Consequently, the Fermi vector and the Fermi energy increase and the Fermi sphere expanded. Energy levels move on an energy scale following density of states reduction. To simplify the presentation, in Fig. 1 and the following related figures, we presume that the energy levels did not change position and are equidistant in the reference well.

In RQW the total energy of the electrons is increased with respect to reference QW. The electron gas in RQW is an excited system. If there was some external mechanism to allow back the forbidden quantum states (QS), then the electrons would occupy them and $E_F^{(R)}$ would get decreased (to minimize the energy of the system).

**3. Mechanism of electron-electron attraction**
We consider tunneling to another RQW as a possible mechanism of energy minimization. Fig.2 shows two RQW placed close enough for tunneling and corresponding energy diagram. Adjacent RQW changes the boundary condition for the electron wave function (non-zero value becomes allowed outside the well). Modification of the boundary condition re-establishes the forbidden QSs (in the limit of zero-width gap between RQWs). The density of the QSs increases back and the Fermi energy decreases back (Fermi sphere shrinks back).

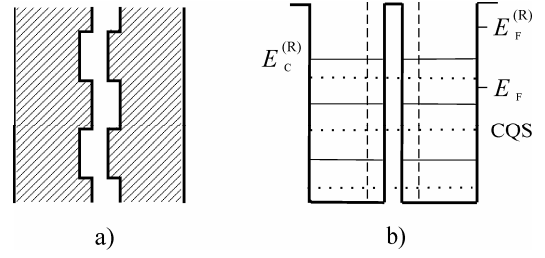

Fig. 2 Two RQWs placed close enough for electron tunneling and the quantum energy levels. Dotted lines depict the correlated quantum states.

Adding another RQW reduces the total energy of the electron gas. Closer it is placed; the higher is the probability of tunneling. The probability of the electron being in the re-established QS increases with the decreasing distance between the wells. The adjacent wells tend to collapse the gap (to reduce the system energy as much as possible). This corresponds to the attractive force. Tunneling occurs in both the directions (because of symmetry) and hence re-established QS cannot be ascribed to a particular RQW (dotted line in Fig. 2b). It belongs to the system of two. Let us name re-established QS as a correlated quantum state (CQS). The probability of occupation of CQS is equal to the tunneling probability.

For more clarity, an additional description of attractive force is given. We divide conventional QW into equal parts in two different ways (cross-sections shown in Fig. 3). First, as shown in Fig. 3a, it is divided by plane, resulting in two conventional

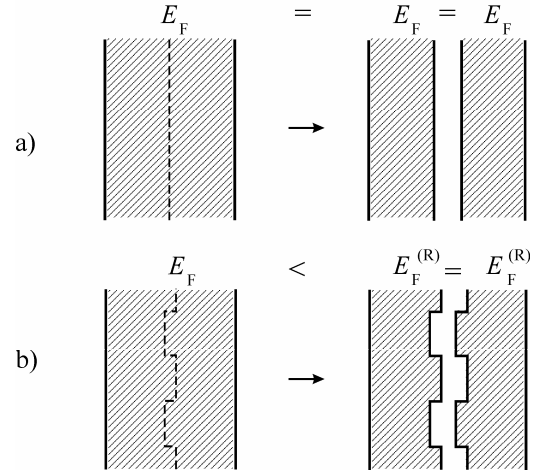

Fig. 3 Two ways of well splitting: a) parts do not attract each other; b) parts attract each other.

QWs. Both wells contain half of the initial number of atoms and free electrons. The Fermi energy of two parts is equal and do not differ from the Fermi energy of the initial well. Separation does not change the energy per electron and there is no attraction force between the parts. Subsequently, we can divide the same

QW by ridged plane, as shown in Fig. 3b. Here, the Fermi energy of both parts increases in the process of division, $E_F^{(R)} > E_F$. Furthermore, energy per free electron in both the RQWs increases. Now, the parts attract each other to retain the initial unity and reduce the system energy. The only difference between the final states in Fig. 3a and Fig. 3b is the electron-gas energy spectrum, and hence, the attraction force originates from it.

Consider that the electron in the CQS is tunneling from left to right RQW. When the electron is inside the barrier (Fig. 4), its positive images [14] emerge in both the RQWs. The electron image is the mathematical representation of the "transport electrons" redistribution inside the well ("transport electrons" are those with energies $E \approx E_F^C \pm 3K_B T$). As electron passes through the barrier, the right image approaches it and the left image moves away from it, and both the images attract the electron. Potentially, the right image can attract one more electron from the right RQW. Thus, the image can potentially serve as a mediator between two electrons and attract them to each other (like positively charged atom centre in BCS theory). Yet, this is not possible under conventional conditions. Electrons from the right well create image thethemselves, and obviously, the image cannot attract its own source. In the case of CQS electrons, the situation gets principally different. The electrons being in the CQS are those with energies $E << E_F^{(C)}$ and, therefore, do not participate in image formation. Usually, they do not participate in charge transport (as all the QSs nearby would already be occupied and the exchange of a small amount of energy with the environment is quantum-mechanically forbidden). Hence, another electron being in CQS can be attracted by the image. Thus, the proposed electron–electron attraction mechanism is as follows: CQS electron with wave vector **k**, attracts its right image via "transport electrons" in the process of tunneling. Right image itself attracts another CQS electron from the right RQW having wave vector, −**k**. As a result, the attraction between two CQS electrons (one being inside the barrier and the other being inside the right RQW) takes place. Since the electron image is only the mathematical representation of the "transport electron" redistribution in space, the real mediator between paired CQS electrons is a collective movement of those "transport electrons".

Described electron–electron attraction could not take place in the system of two conventional QWs. First, it will not work for electrons with $E << E_F^{(C)}$, since all QSs in both the wells are already occupied in that energy range (tunneling requires empty QS in the receiving QW). Second, it will not work for electrons with energies, $E \approx E_F^{(C)} \pm 3K_B T$ and $E >> E_F^{(C)}$ (both having empty QSs around), since these electrons participate in the image formation themselves.

Electron being in the CQS can have wave vectors **k** and –**k** (tunneling from left to right or in opposite direction). Consequently, there are four possible QSs, **k**↑, **k**↓, −**k**↑ and −**k**↓. Utmost, four inter-tunneling pairs, **k**↑ −**k**↑, **k**↑ −**k**↓, **k**↓ −**k**↑, and **k**↓ -**k**↓ could be obtained from them. However, we exclude the first and last ones, since electrons with the same spin cannot be placed close in real space, and the remaining two are Cooper pairs. Therefore, CQS occupied by a maximum of two Cooper pairs, **k**↑ −**k**↓ and **k**↓ − **k**↑ can re-establish to reduce the total energy of the system. The pairs and generally CQS do not remain stationary, since the tunneling probability is low. On the other hand, the density of CQSs is high and the product results in some finite number of Cooper pairs existing at the same time.

## 4. Cooper-pairs in cuprates and the pseudogap

In cuprates, O and Cu atoms are shifted up and down, relative to the common plane of the $CuO_2$ layer. The geometry of the layer is akin to the periodic ridges of RQW. Although the $CuO_2$ layer has no firm boundaries, it is evident that its boundaries are not planar. The boundaries do have some geometry, even in the Hg-based cuprates, where the centers of Cu and O atoms are exactly in the same plane. [15] However, the geometry exists owing to different radii of Cu and O atoms (ions). Hence, in the first approximation, we regard a $CuO_2$ layer as an RQW-containing electron gas and the layer have forbidden QSs and expanded Fermi sphere. The FSE forces the electron gases in the adjacent layers to reduce their total energy, by means of CQSs.

To verify the model we calculate the reduction of energy per electron in the system of two $CuO_2$ layers and compare it with measured pseudogap values. Fig. 5 shows two

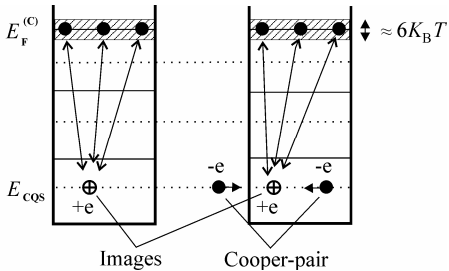

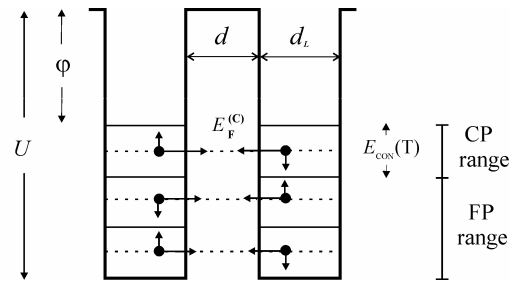

Fig. 5 The CQS (dotted lines) occupied by Cooper pairs.

$CuO_2$ layers separated by distance $d$. The density of QSs as found in [7] is

$$\rho_{RQW}(E) = \rho_0(E)/G . \quad (1)$$

Fig. 4 CQS electron in the process of tunneling between two RQW and image-mediated electron–electron attraction.

Here, $\rho_{RQW}(E)$ is the density of QSs in RQW, $\rho_0(E)$ is the density of QSs in conventional QW, $E$ is the electron energy, and $G$ is the geometry factor (Fig. 1). Thus, the density of forbidden QSs is

$$\rho^-(E) = \rho_0(E) - \rho_{RQW}(E) = \rho_0(E)(1 - G^{-1}). \quad (2)$$

Each electron that leaves the Fermi level to CQS can be in four possible QSs, $\mathbf{k}\uparrow$, $\mathbf{k}\downarrow$, $-\mathbf{k}\uparrow$, and $-\mathbf{k}\downarrow$. Therefore, the CQS energy-level degeneracy is four. The tunneling probability can be written as

$$D(E) = \exp\left[-\frac{2d}{\hbar}\sqrt{2m(U-E)}\right]. \quad (3)$$

Where, $\hbar$ is the Plank's constant, $m$ is the electron mass, and $U$ is the height of the potential barrier. The probability of electron being in CQS is equal to the tunneling probability. In this study, it has been assumed that the receiving QS is almost empty, i.e. $1 - D(E) \approx 1$, as $D(E) \ll 1$. "Diving" the electron into CQS leads to energy reduction and binding of adjacent layers. Thus, the layer binding energy density (per unit volume) within the energy interval of $\delta E$ will be

$$\delta E_{bin} = 4\varepsilon_{CQS} D(E) \rho^-(E) \delta E \quad (4)$$

where $E_{bin}$ is the layer binding energy (per unit volume) and $\varepsilon_{CQS} = (U - \varphi - E)$ is the reduction of energy per electron in the process of CQS formation. The factor four comes from level degeneracy. Integration of Eq. (4) over the energy range below the Fermi energy gives

$$E_{bin} = 4\int_{U-\varphi}^{0} (U - \varphi - E) D(E) \rho^-(E) dE \quad (5)$$

Inserting Eq. (2) in Eq. (5) results in

$$E_{bin} = 4(1 - G^{-1})\int_{U-\varphi}^{0} (U - \varphi - E) D(E) \rho_0(E) dE \quad (6)$$

Equation (6) contains the density of QSs, $\rho_0(E)$, for the conventional QW, and the well-known formula for 3D quantum well, $\rho_0(E) = m\sqrt{2mE}/\pi^2\hbar^3$, have been used. Finally, the layer binding energy density can be written as

$$E_{bin} = \frac{4\sqrt{2}\,m^{3/2}(1 - G^{-1})}{\pi^2\hbar^3} \times$$
$$\times \int_{U-\varphi}^{0} (U - \varphi - E) D(E) \sqrt{E}\, dE. \quad (7)$$

**5. The binding energy per electron and comparison with pseudogap values**
To obtain the binding energy per electron, we divide $E_{bin}$ by the density of the electrons. In hole-doped cuprates, the density of electrons in $CuO_2$ layers is equal to the density of the holes in charge reservoirs, and has the universal value [16] for optimally doped cuprates, $p=1.6\times10^{21}$ cm$^{-3}$. Consequently, the binding energy per electron in such cuprates is $\varepsilon_{bin} = E_{bin}/p$. Let us calculate the values of $\varepsilon_{bin}$ for some double-layer cuprates and compare it with the measured pseudogap values. Further, it is assumed that $G \gg 1$ and $(1 - G^{-1}) \approx 1$ (strictly, $G$ depends on buckling angle of $CuO_2$ layer, but it can be ignored in first approximation). The experimental values of interlayer distance for the two-layer cuprates were 3.36 Å for YBCO, 3.35 Å for Bi2212, 3.2 Å for Tl2212 (all three from Ref. 17), and 3.23 Å for Hg1212 (Ref. 15). The listed values are the distances between the atom centers and include the dimensions of the electron clouds. Let the electron cloud radius in tunneling direction be $R_c$. Subsequently, we subtract $2R_c$ from the interlayer distance. Thus, we get

$$\varepsilon_{bin} = \frac{4\sqrt{2}\,m^{3/2}}{\pi^2\hbar^3 p} \int_{U-\varphi}^{0} (U - \varphi - E)\sqrt{E}\, \times$$
$$\times \exp\left[-\frac{2(d - 2R_c)}{\hbar}\sqrt{2m(U-E)}\right] dE. \quad (8)$$

The following experimental values were inserted in Eq. (8): for a work function, [18] $\varphi = 4$ eV; for Fermi energy, [19, 20] $E_F^{(C)} = U - \varphi = 300$ meV, and for interlayer distance, $d=3.2$ Å. The Cu and O atoms have the atomic radii of 1.28 Å and 0.73 Å, respectively, and the ionic radii [21] are 0.87 Å for $Cu^{2+}$ and 1.26 Å for $O^{2-}$. However, it is not clear on which value should be used for effective $R_c$. The natural suggestion is that it should be in the range of 0.73 Å $< R_c <$ 1.26 Å. Fig. 6 shows the plot of $\varepsilon_{bin}$ as the function of $2R_c$, according to Eq. (8), in the above mentioned range of $R_c$. The figure also shows the experimental values of pseudo gaps obtained from the tunneling spectroscopy [22-26]. The values calculated within our model fit the experimental ones in the reasonable range of $R_c$.

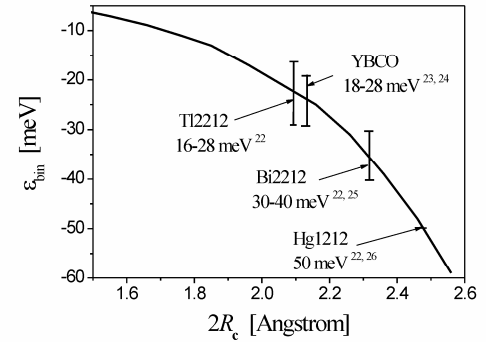

Fig. 6 Solid curve corresponds to binding energy per electron, $\varepsilon_{bin}$. The experimental values of pseudogap for optimally doped two-layer cuprates are given for comparison.

It is essential to compare the electron binding energy in a Cooper pair with the energy reduction (per electron) during CQS formation. The Coulomb attrac-tion between the paired electrons (with positive image +e in the core), gives the binding energy per electron, $(e^2/4\pi\varepsilon_0)(3/4d_L) = 5\,\text{eV}$, for the layer thickness, $d_L = 2$ Å (screening is neglected, since the layer thickness $d_L$ is only 1–2 atom size). In addition, the maximum energy reduction during the transition of electron from the Fermi level to CQS is only $E_F^{(C)} \approx 0.3\,\text{eV}$ in cuprates. Consequently, the image-mediated electron–electron attraction can easily provide the needed energy reduction. It is interesting to note that the latter is not applicable for conventional solids having $E_F \approx 10$ eV.

The Fermi energy in cuprates is low and the corresponding de Broglie wavelength $\lambda_F \propto 1/\sqrt{E_F}$ is high. The relatively large $\lambda > \lambda_F = 15\text{–}25$ Å allows the electron to participate in the tunneling events at long distances, possibly as far as three $CuO_2$ layers.[27, 28] Therefore, tunneling between three or more layers might contribute significantly. Multi-layer tunneling increases the binding of the layers and allows further reduction in the total energy of the system. This may explain the increase in $T_c$ with the increasing number of similar $CuO_2$ layers per unit cell from 1 to 3.

The described model shows good quantitative agreement with the experiment in the case of multi-layer cuprates having $d=3\text{–}4$ Å. However, in the case of single-layer cuprates, $d=6\text{–}12$ Å and $\varepsilon_{bin}$ becomes <0.1 meV. This value is much lesser than the experimental pseudogap values. A possible reason for the high pseudogap value in the single-layer cuprates is the negative-U centers inside the charge reservoir layers. They can serve as resonant tunneling centers and increase the tunneling probability [29] by reducing the effective distance between the layers down to 3–4 Å.

## 6. Bose condensation

In conventional superconductors, the electron–phonon interaction is responsible for Cooper-pair formation (BCS theory), and the binding energy per electron is of the order of 1 meV. Paired electrons have wave vectors close to Fermi wave vector $\mathbf{k} \approx \mathbf{k}_F$. In the presented mechanism, the pair-binding energy (per electron) is $\Delta_{CQS}(\mathbf{k}) = E_F^{(R)} - E_{CQS}(\mathbf{k})$ and vary in a wide range of 0–300 meV ($E_F^{(C)} \approx 300$ meV in cuprates) as $|\mathbf{k}|$ varying considerably from pair to pair. The pair-effective mass, $M$, being proportional to $\Delta_{CQS}(\mathbf{k})$ according to negative-U Hubbard model,[30] also vary in a wide range. Therefore, the pairs have very different starting conditions for phase ordering and Bose condensation. With decreasing $T$, the pairs having low $M$ will Bose condense prior to those having high $M$. This can explain the two energy gaps and the two characteristic temperatures in cuprates. Condensed pairs (CP) result in superconductive gap $\Delta_C$ and free pairs (FP) result in pseudogap $\Delta_P$. As FPs have more $\Delta_{CQS}(\mathbf{k})$, $\Delta_P > \Delta_C$. The FP and $\Delta_P$ exist below $T^*$, while CP and $\Delta_C$ exist only below $T_C$.

Let $E_{CON}(T)$ be the maximum $\Delta_{CQS}(\mathbf{k})$ that allows Bose condensation at a given $T$. The energy interval $E < E_F^{(C)}$ can be formally divided into two regions (Fig. 5), namely the CP region, where $\Delta_{CON}(\mathbf{k}) < E_{CON}(T)$ and FP region, where $\Delta_{CON}(\mathbf{k}) > E_{CON}(T)$. At $T=0$, all the pairs are CP, within the whole range $E < E_F^{(C)}$. There would be no region of FP, since at $T=0$ all pairs condense independent of $M$. When $T$ is increased, some pairs leave the condensate owing to high $M$, and FP with $\Delta_{CON}(\mathbf{k}) \leq E_F^{(C)}$ emerge. With further increase in $T$, the CP region shrinks and disappears at $T=T_C$. However, the FP region remains above $T_C$. A further increase in $T$ reduces the number of FP owing to thermal fluctuations, and all pairs get destroyed at $T^*$.

Both $E_{CON}(T)$ and $T_C$ depends on the phase-ordering mechanism, which is out of scope of this study. Still, we make one general note. Strong layer binding corresponds to more order and less entropy $S$ of the system. The layer binding energy has not only the CP component, but also the FP component. Consequently, the free pairs influence condensation process indirectly. They increase the layer binding and reduce $S$.

## 7. Fermi surface pockets as a consequence of layer geometry

The Fermi surface pockets were found in APRES and Shubnikov–de Haas effect measurements. Fermi surface area is significantly reduced in the pseudogap (or under-doped) regime. Our model provides a possible explanation for Fermi surface pockets. Geometry of $CuO_2$ layer modifies the Fermi surface area and shape. The $\mathbf{k}$ spectrum in the ridged geometry has been investigated earlier.[9] However, we will underline some related details here. Fig. 7a shows the ridged well and the corresponding $\mathbf{k}$ spectrum in $\mathbf{k}_y$, $\mathbf{k}_z$ plane ($a,b$ plane in cuprates). Electrons having low wave vector

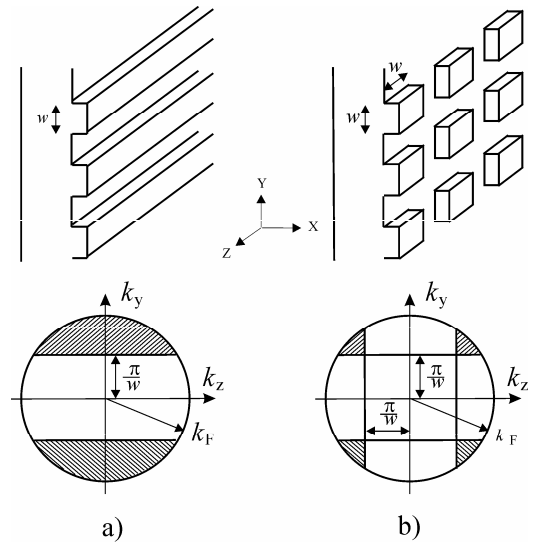

Fig. 7 a). RQW and its $k_y$, $k_z$ spectrum. b) Modified RQW and its $k_y$, $k_z$ spectrum akin to Fermi surface pockets in cuprates.

component $|\mathbf{k}_y| < \pi/w$ cannot exist in the well of such geometry (such waves cannot "fit" inside the ridges). Thus, owing to ridges, special boundary conditions on electron-wave function is imposed. Simultaneously, **k** plane contains an external circle of diameter $\mathbf{k}_F$ (maximum possible k at $T$=0). The circle and the two lines $k_y = \pm \pi/w$ limit the allowed **k** area, shown as the shaded portion at the bottom of Fig. 7a.

Subsequently, we replace the ridges by right-square prisms as shown in Fig. 7b. Here, the Z component of the wave vector $\mathbf{k}_z$ also gets filtered and $|\mathbf{k}_z| < \pi/w$ becomes forbidden as well. As a result, we get four allowed **k** areas, represented by the shaded portion in the bottom of Fig. 7b. The allowed **k** areas in Fig. 7b are akin to Fermi surface pockets observed in cuprates. The prisms represent Cu atoms. Obviously, the geometry of Cu-atom electron cloud differs from the prisms and is more like a dome. Thus, the shape of Fermi surface pockets should also differ from those shown in Fig. 7(b). Still, the model explains (quantitatively) the existence of Fermi surface pockets in cuprates.

More accurate results can be obtained by using special mathematical methods recently developed for Casimir energy calculation. Wave spectrum inside the vacuum gap exhibits strong dependence on gap geometry.[31] The number of geometries, including the double-side ridged geometry [32] and the double-side corrugated geometry [33] were also analyzed.

## 8. Comparison with experiments

The possible model of pairing described in this study successfully explains the following cuprate properties:

(i) Loss of superconductivity in ultra-thin films – The $T_c$ reduces when the film thickness decreases, and the superconductivity vanishes after less than two $CuO_2$ layers are left.[7] In the described model, it was observed that at least two $CuO_2$ layers should be present to allow Cooper pairing and superconductivity.

(ii) Pseudogap is present in the energy spectrum above $T_c$, and its width does not depend on $T$ – In the model described, the pseudogap is formed by FP. The pseudogap width does not depend on $T$, since it is a consequence of free pairs, having pairing energy up to 300 meV, which is much more than $K_B T^* \approx 10$ meV.

(iii) Electronic specific heat, $C_v$ – At $T^*$, the ratio $C_v/T$ starts to reduce with the decrease in $T$ (Ref. 20, 34). It changes behavior at $T^*$, and not at $T_c$ (as in conventional superconductors). Our model provides a possible explanation. When $T$ is decreased, FPs start forming at $T^*$, and they thermodynamically decouple from the electron gas. Consequently, the electronic specific heat decreases, starting from $T^*$.

(iv) Scaling relationship between $T_c$ and the buckling angle of the $CuO_2$ planes [35, 36] – The buckling angle sets the layer geometry and consequently, the value of $G$ and the layer binding energy. Earlier experiments revealed that phonons are not involved in Cooper pair formation. Consequently, it became very difficult to explain the buckling-angle (or internal and external pressure) dependence of $T_c$; however, the model described in this study provides a natural explanation through geometry changes.

(v) Anisotropy of electrical-conductivity $\sigma(T)$ dependence in the pseudogap phase – Earlier experiments demonstrated that it is semiconductor-like $(d\sigma/dT)_c < 0$ in $c$ direction and metal-like $(d\sigma/dT)_{ab} > 0$ in $ab$ plane. In the presented model, the electrons with $\mathbf{k} \approx \mathbf{k}_F$ have large anisotropy in **k**, and those with large $\mathbf{k}_c$ participate in the formation of CQSs. Consequently, the electrons with high $\mathbf{k}_c$ are absent in the **k** spectrum near $\mathbf{k} \approx \mathbf{k}_F$. Such electrons "dive" from the region $\mathbf{k} \approx \mathbf{k}_F$ into CQS and form Cooper pairs. Empty $\mathbf{k}_c \approx \mathbf{k}_F$ region results in semiconductor-like behavior of $\rho(T)$ in $c$ direction. On the other hand, the electrons with $\mathbf{k} \approx \mathbf{k}_F$ and low $\mathbf{k}_c$ (high $\mathbf{k}_a$ or $\mathbf{k}_b$) do not participate in the formation of CQSs and, therefore, $\rho(T)$ dependence in $ab$ plane remains metal-like. The scaling of $c$-axis resistivity with pseudogap energy[37] is in full agreement with the present model.

(vi) ARPES data is collected only for $\mathbf{k}_a$ and $\mathbf{k}_b$, and $\mathbf{k}_c$, being normal to surface component, is not measured (general problem of photoemission spectroscopy[4]), since the cuprate crystals are cleaved in situ along the $CuO_2$ plane. Consequently, the information about $\mathbf{k}_c$ is absent in APRES data.[6, 38] According to the described model, the electron pairing introduces changes exactly in $\mathbf{k}_c$. This explains why a large number of precise ARPES data are unable to reveal the pairing mechanism so far.

(vii) Recently investigated iron-based HTS materials also have layered structure with conducting AsFe layer. The AsFe layer has the geometry close to the ridged-like $CuO_2$ layer of cuprates.

(viii) Superconductivity emerges at the interface of insulator and metal materials. [2, 8] This can be explained by a specific geometry of $CuO_2$ layers from both sides and electronic structure. Possible scenario is that layer geometry is suitable for CQS-formation in insulating material but there are not enough electrons in it to form Cooper pairs. The metal interface serves as an electron source. Tunneling between the two layers on opposite sides of the interface allows CQSs and Copper pairs. This scenario does not explain why the second (counting from the interface) $CuO_2$ play a major role in the superconductivity and first layer does not.[8] However, in experiments, first layer may have damaged geometry owing to impurities or surface tension. This provides a possible explanation for independence of $T_c$ on doping of the first layer.

## 9. Conclusions

In this study, the possible mechanism of electron–electron attraction in cuprates, based on image force, is proposed. Electrons tunnel between adjacent $CuO_2$ layers to reduce the energy of the system. Initially, owing to the Fermi sphere expansion, the energy of electron gas possesses added value in the individual layers. Electron tunneling allows the formation of CQS, resulting in reduction of system energy, and the Cooper pairs occupy the CQS. At $T>0$, depending on their individual binding energy, some pairs Bose condense, while the others remain free. The condensed pairs are responsible for superconductive gap and free pairs are responsible for pseudogap. The energy reduction per electron, calculated within the model, is in agreement with the experimental values of pseudogap for the two-layer cuprates, such as YBCO, Bi2212, Tl2212, and Hg1212. The possible model described in this study explains the two energy gaps and the two characteristic temperatures in cuprates. It also provides an explanation to the low-order parameter, Fermi surface pockets, unconventional isotopic effect, conductance anisotropy in the pseudogap state, temperature dependence of electronic specific heat, vanishing of superconductivity in ultra-thin films, scaling relationship between buckling angle and $T_c$, and other properties of HTS cuprates. The presented mechanism provides a possible explanation of superconductivity in other layered materials and structures such as iron based superconductors and LSCO/LCO interfaces.

**Acknowledgments**
I am grateful to G. Gabadadze for useful discussions. This work was partly sponsored by Borealis Technical Limited.